\documentclass[showkeys,showpacs,notitlepage]{revtex4-1}%
\usepackage{amsmath,amssymb,amsthm,graphicx,slashed}%
\usepackage{amsmath}%
\usepackage{amsfonts}%
\usepackage{amssymb}%
\usepackage{graphicx}
\providecommand{\U}[1]{\protect\rule{.1in}{.1in}}
\pdfoutput=1

\DeclareMathOperator{\GeV}{GeV}

\def\U{\mathcal{U}}
\begin{document}
\title[Grand Unification in the Spectral Pati--Salam]{Grand Unification in the Spectral Pati--Salam Model}
\author{Ali H. Chamseddine$^{1,3}$, Alain Connes$^{2,3,4}$ and Walter D. van
Suijlekom$^{5}$}
\email{chams@aub.edu.lb, alain@connes.org, waltervs@math.ru.nl}
\affiliation{$^{1}$Physics Department, American University of Beirut, Lebanon}
\affiliation{$^{2}$College de France, 3 rue Ulm, F75005, Paris, France}
\affiliation{$^{3}$I.H.E.S. F-91440 Bures-sur-Yvette, France}
\affiliation{$^{4}$Department of Mathematics, The Ohio State University, Columbus OH 43210 USA}
\affiliation{$^{5}$Institute for Mathematics, Astrophysics and Particle Physics, Radboud
University Nijmegen, Heyendaalseweg 135, 6525 AJ Nijmegen, The Netherlands}

\begin{abstract}
We analyze the running at one-loop of the gauge couplings in the spectral
Pati--Salam model that was derived in the framework of noncommutative geometry.
There are a few different scenario's for the scalar particle content which are
determined by the precise form of the Dirac operator for the finite
noncommutative space. We consider these different scenarios and establish for
all of them unification of the Pati--Salam gauge couplings. The boundary
conditions are set by the usual RG flow for the Standard Model couplings at an
intermediate mass scale at which the Pati--Salam symmetry is broken.

\end{abstract}
\maketitle

\section{Introduction}

This paper builds on two recent discoveries in the noncommutative geometry
approach to particle physics: we showed in \cite{CCS13b} how to obtain inner
fluctuations of the metric without having to assume the order one condition on
the Dirac operator. Moreover the original argument by classification
\cite{CC07b} of finite geometries $F$ that can provide the fine structure of
Euclidean space-time as a product $M\times F$ (where $M$ is a usual
$4$-dimensional Riemannian space) has now been replaced by a much stronger
uniqueness statement \cite{CCM14,CCM15}. This new result shows that the
algebra
\begin{equation}
M_{2}(\mathbb{H})\oplus M_{4}(\mathbb{C}),\label{coor}%
\end{equation}
where $\mathbb{H}$ are the quaternions, appears uniquely when writing the
higher analogue of the Heisenberg commutation relations. This analogue is
written in terms of the basic ingredients of noncommutative geometry where one
takes a spectral point of view, encoding geometry in terms of operators on a
Hilbert space $\mathcal{H}$. In this way, the inverse line element is an
unbounded self-adjoint operator $D$. The operator $D$ is the tensor sum of the
usual Dirac operator on $M$ and a `finite Dirac operator' on $F$, which is
simply a hermitian matrix $D_{F}$. The usual Dirac operator involves $\gamma$
matrices which allow one to combine the momenta into a single operator. The
higher analogue of the Heisenberg relations puts the spatial variables on
similar footing by combining them into a single operator $Y$ using another set
of $\gamma$ matrices and it is in this process that the algebra \eqref{coor}
appears canonically and uniquely in dimension $4$. We refer to
\cite{CCM14,CCM15} for a detailed account. What matters for the present paper
is that the above process leads without arbitrariness to the Pati--Salam
\cite{PS74} gauge group $SU(2)_{R}\times SU(2)_{L}\times SU(4)$, together with
the corresponding gauge fields and a scalar sector, all derived as inner
perturbations of $D$ \cite{CCS13b}. Note that the scalar sector can not be
chosen freely, in contrast to the early work on Pati--Salam unification
\cite{Eli76,Eli77,Bar82,CMP84}. In fact, there are only a few possibilities
for the precise scalar content, depending on the assumptions made on the
finite Dirac operator.

From the spectral action principle, the dynamics and interactions are
described by the \emph{spectral action} \cite{CC96,CC97},
\begin{equation}
\mathrm{tr}(f(D_{A}/\Lambda))
\end{equation}
where $\Lambda$ is a cutoff scale and $f$ an even and positive function. In
the present case, it can be expanded using heat kernel methods,
\begin{equation}
\mathrm{tr}(f(D_{A}/\Lambda))\sim F_{4}\Lambda^{4}a_{0}+F_{2}\Lambda^{2}%
a_{2}+F_{0}a_{4}+\cdots\label{eq:sa}%
\end{equation}
where $F_{4},F_{2},F_{0}$ are coefficients related to the function $f$ and
$a_{k}$ are Seeley deWitt coefficients, expressed in terms of the curvature of
$M$ and (derivatives of) the gauge and scalar fields. This action is
interpreted as an effective field theory for energies lower than $\Lambda$.

One important feature of the spectral action is that it gives the usual
Pati--Salam action with unification of the gauge couplings \cite{CCS13b} (cf.
Eq. \eqref{eq:sa-ps} below). This is very similar to the case of the spectral
Standard Model \cite{CCM07} where there is unification of  gauge couplings.
Since it is well known that the SM gauge couplings do not meet exactly, it is
crucial to investigate the running of the Pati--Salam gauge couplings beyond
the Standard Model and to find a scale $\Lambda$ where there is grand
unification:
\begin{equation}
g_{R}(\Lambda)=g_{L}(\Lambda)=g(\Lambda).
\end{equation}
This would then be the scale at which the spectral action \eqref{eq:sa} is
valid as an effective theory. There is a hierarchy of three energy scales: SM,
an intermediate mass scale $m_{R}$ where symmetry breaking occurs and which is related to the neutrino Majorana masses
($10^{11}-10^{13}$Gev), and the GUT scale $\Lambda$.

For simplicity, we restrict our analysis to the running of the gauge couplings
at one-loop. Indeed, at two loops, the gauge and scalar couplings are mixed
and influence each other. Moreover, the running of the scalar mass terms can
not be trusted at all because of quadratic divergences.

Thus, we analyze the running of the gauge couplings according to the usual
(one-loop) RG equation where each takes the form
\begin{equation}
16\pi^{2}\frac{dg}{dt}=-bg^{3}.\label{eq:RG-general}%
\end{equation}
The coefficient $b$ is determined by the particle content and their
representation theory \cite{CEL74,MV83,MV84,MV85} for which we use \cite{KP81}
as well as the program PyR@TE. As mentioned before, depending on the
assumptions on $D_{F}$, one may vary to a limited extent the scalar particle
content, consisting of either composite or fundamental scalar fields. We will
not limit ourselves to a specific model but consider all cases separately. In
fact, we establish grand unification for all of them, thus confirming validity
of the spectral action at the corresponding scale, independent of the specific
form of $D_{F}$.

\section{Spectral Pati--Salam and grand unification}

One of the pressing questions at present is whether there is new physics
beyond the Standard Model. The success of the spectral construction of the
Standard Model, predicting its particle content, including gauge fields, Higgs
fields as well as a singlet whose vev gives Majorana mass to the right handed
neutrino, is a strong signal that we are on the right track. The route that
led to this conclusion starts with classifying the algebras of the finite
space. The results show that the only algebras which solve the fermion
doubling problem are of the form $M_{2a}(\mathbb{C})\oplus M_{2a}(\mathbb{C})$
where $a$ is an even integer. An arbitrary symplectic constraint is imposed on
the first algebra restricting it from $M_{2a}(\mathbb{C})$ to $M_{a}%
(\mathbb{H}).$ The first non-trivial algebra one can consider is for $a=2$
with the algebra \cite{CC07b}
\begin{equation}
M_{2}(\mathbb{H})\oplus M_{4}(\mathbb{C}).
\end{equation}
Coincidentally, and as explained in the introduction, the above algebra comes
out as a solution of the two-sided Heisenberg quantization relation between
the Dirac operator $D$ and the two maps from the four spin-manifold and the
two four spheres $S^{4}\times S^{4}$ \cite{CCM14,CCM15}. This removes the
arbitrary symplectic constraint and replaces it with a relation that quantize
the four-volume in terms of two quanta of geometry and have far reaching
consequences on the structure of space-time.

The existence of the chirality operator $\gamma$ that commutes with the
algebra breaks the quaternionic matrices $M_{2}(\mathbb{H})$ to the diagonal
subalgebra and leads us to consider the finite algebra
\begin{equation}
\mathcal{A}_{F}=\mathbb{H}_{R}\oplus\mathbb{H}_{L}\oplus M_{4}(\mathbb{C}).
\end{equation}
The Pati--Salam gauge group $SU(2)_{R}\times SU(2)_{L}\times SU(4)$ is
obtained as the inner automorphism group of $\mathcal{A=C}^{\infty}\left(
M\right)  \otimes\mathcal{A}_{F}$, and the corresponding gauge bosons appear
as inner perturbations of the (spacetime) Dirac operator \cite{CCS13b}.

Next, an element of the Hilbert space $\Psi\in\mathcal{H}$ is represented by
\begin{equation}
\Psi_{M}=\left(
\begin{array}
[c]{c}%
\psi_{A}\\
\psi_{A^{^{\prime}}}%
\end{array}
\right)  ,\quad\psi_{A^{\prime}}=\psi_{A}^{c}%
\end{equation}
where $\psi_{A}^{c}$ is the conjugate spinor to $\psi_{A}.$ Thus all primed
indices $A^{\prime}$ correspond to the Hilbert space of conjugate spinors. It
is acted on by both the left algebra $M_{2}\left(  \mathbb{H}\right)  $ and
the right algebra $M_{4}\left(  \mathbb{C}\right)  $. Therefore the index $A$
can take $16$ values and is represented by
\begin{equation}
A=\alpha I
\end{equation}
where the index $\alpha$ is acted on by quaternionic matrices and the index
$I$ \ by $M_{4}\left(  \mathbb{C}\right)  $ matrices. Moreover, when the
grading breaks $M_{2}\left(  \mathbb{H}\right)  $ into $\mathbb{H}_{R}%
\oplus\mathbb{H}_{L}$ the index $\alpha$ is decomposed to $\alpha
=\overset{.}{a},a$ where $\overset{.}{a}=\overset{.}{1},\overset{.}{2}$
(dotted index) is acted on by the first quaternionic algebra $\ \mathbb{H}%
_{R}$ and $a=1,2$ is acted on by the second quaternionic algebra
$\ \mathbb{H}_{L}$. When $M_{4}\left(  \mathbb{C}\right)  $ breaks into
$\mathbb{C}\oplus M_{3}\left(  \mathbb{C}\right)  $ (due to symmetry breaking
or through the use of the order one condition as in \cite{CC07b}) the index
$I$ is decomposed into $I=1,i$ and thus distinguishing leptons and quarks,
where the $1$ is acted on by the $\mathbb{C}$ and the $i$ by $M_{3}\left(
\mathbb{C}\right)  .$ Therefore the various components of the spinor $\psi
_{A}$ are
\begin{align}
\psi_{\alpha I} &  =\left(
\begin{array}
[c]{cccc}%
\nu_{R} & u_{iR} & \nu_{L} & u_{iL}\\
e_{R} & d_{iR} & e_{L} & d_{iL}%
\end{array}
\right)  ,\qquad i=1,2,3\\
&  =\left(  \psi_{\overset{.}{a}1},\psi_{\overset{.}{a}i},\psi_{a1},\psi
_{ai}\right)  ,\qquad a=1,2,\quad\overset{.}{a}=\overset{.}{1},\overset{.}{2}%
\nonumber
\end{align}
This is a general prediction of the spectral construction that there is $16$
fundamental Weyl fermions per family, $4$ leptons and $12$ quarks.

The (finite) Dirac operator can be written in matrix form%
\begin{equation}
D_{F}=\left(
\begin{array}
[c]{cc}%
D_{A}^{B} & D_{A}^{B^{^{\prime}}}\\
D_{A^{^{\prime}}}^{B} & D_{A^{^{\prime}}}^{B^{^{\prime}}}%
\end{array}
\right)  ,\label{eq:dirac}%
\end{equation}
and must satisfy the properties
\begin{equation}
\gamma_{F}D_{F}=-D_{F}\gamma_{F}\qquad J_{F}D_{F}=D_{F}J_{F}%
\end{equation}
where $J_{F}^{2}=1.$ A\ matrix realization of $\gamma_{F}$ and $J_{F}$ are
given by
\begin{equation}
\gamma_{F}=\left(
\begin{array}
[c]{cc}%
G_{F} & 0\\
0 & -\overline{G}_{F}%
\end{array}
\right)  ,\qquad G_{F}=\left(
\begin{array}
[c]{cc}%
1_{2} & 0\\
0 & -1_{2}%
\end{array}
\right)  ,\qquad J_{F}=\left(
\begin{array}
[c]{cc}%
0_{4} & 1_{4}\\
1_{4} & 0_{4}%
\end{array}
\right)  \circ\mathrm{cc}%
\end{equation}
where $\mathrm{cc}$ stands for complex conjugation. These relations, together
with the hermiticity of $D$ imply the relations
\begin{equation}
\left(  D_{F}\right)  _{A^{^{\prime}}}^{B^{^{\prime}}}=\left(  \overline
{D}_{F}\right)  _{A}^{B}\,\qquad\left(  D_{F}\right)  _{A^{^{\prime}}}%
^{B}=\left(  \overline{D}_{F}\right)  _{B}^{A^{\prime}}%
\end{equation}
and have the following zero components \cite{CC10}
\begin{align}
\left(  D_{F}\right)  _{aI}^{bJ} &  =0=\left(  D_{F}\right)  _{\overset{.}{a}%
I}^{\overset{.}{b}J}\\
\left(  D_{F}\right)  _{aI}^{\overset{.}{b}^{\prime}J^{\prime}} &  =0=\left(
D_{F}\right)  _{\overset{.}{a}I}^{b^{\prime}J\prime}%
\end{align}
leaving the components $\left(  D_{F}\right)  _{aI}^{\overset{.}{b}J}$,
$\left(  D_{F}\right)  _{aI}^{b^{\prime}J^{\prime}}$ and $\left(
D_{F}\right)  _{\overset{.}{a}I}^{\overset{.}{b}^{\prime}J^{\prime}}$
arbitrary. These restrictions lead to important constraints on the structure
of the connection that appears in the inner fluctuations of the Dirac
operator. In particular the operator $D$ of the full noncommutative space
given by
\begin{equation}
D=D_{M}\otimes1+\gamma_{5}\otimes D_{F}%
\end{equation}
gets modified to
\begin{equation}
D_{A}=D+A_{\left(  1\right)  }+JA_{\left(  1\right)  }J^{-1}+A_{\left(
2\right)  }%
\end{equation}
where
\begin{equation}
A_{\left(  1\right)  }=%
{\displaystyle\sum}
a\left[  D,b\right]  ,\,\qquad A_{2}=%
{\displaystyle\sum}
\widehat{a}\left[  A_{\left(  1\right)  },\widehat{b}\right]  ,\qquad
\widehat{a}=JaJ^{-1}%
\end{equation}

We have shown in \cite{CCS13b} that components of the connection $A$ which are
tensored with the Clifford gamma matrices $\gamma^{\mu}$ are the gauge fields
of the Pati--Salam model with the symmetry of $SU\left(  2\right)  _{R}\times
SU\left(  2\right)  _{L}\times SU\left(  4\right)  .$ On the other hand, the
non-vanishing components of the connection which are tensored with the gamma
matrix $\gamma_{5}$ are given by
\begin{equation}
\left(  A\right)  _{aI}^{\overset{.}{b}J}\equiv\gamma_{5}
\Sigma _{aI}^{\overset{.}{b}J},\qquad\left(  A\right)  _{aI}%
^{b^{\prime}J^{\prime}}=\gamma_{5}H_{aIbJ},\qquad\left(  A\right)
_{\overset{.}{a}I}^{\overset{.}{b}^{\prime}J^{\prime}}\equiv\gamma
_{5}H_{\overset{.}{a}I\overset{.}{b}J}%
\end{equation}
where $H_{aIbJ}=H_{bJaI}$ and $H_{\overset{.}{a}I\overset{.}{b}J}%
=H_{\overset{.}{b}J\overset{.}{a}I}$, which is the most general Higgs
structure possible. These correspond to the representations with respect to
$SU\left(  2\right)  _{R}\times SU\left(  2\right)  _{L}\times SU\left(
4\right)  :$%
\begin{align}
\Sigma_{aI}^{\overset{.}{b}J} &  =\left(  2_{R},2_{L},1\right)  +\left(
2_{R},2_{L},15\right)  \\
H_{aIbJ} &  =\left(  1_{R},1_{L},6\right)  +\left(  1_{R},3_{L},10\right)  \\
H_{\overset{.}{a}I\overset{.}{b}J} &  =\left(  1_{R},1_{L},6\right)  +\left(
3_{R},1_{L},10\right)
\end{align}
We note, however, that the inner fluctuations form a semi-group and if a
component $\left(  D_{F}\right)  _{aI}^{\overset{.}{b}J}$ or $\left(
D_{F}\right)  _{aI}^{b^{\prime}J^{\prime}}$ or $\left(  D_{F}\right)
_{\overset{.}{a}I}^{\overset{.}{b}^{\prime}J^{\prime}}$ vanish, then the
corresponding $A$ field will also vanish. We distinguish three cases: 1)
Left-right symmetric Pati--Salam model with fundamental Higgs fields
$\Sigma_{aI}^{\overset{.}{b}J},$ $H_{aIbJ}$ and $H_{\overset{.}{a}%
I\overset{.}{b}J}.$ In this model the field $H_{aIbJ}$ should have a zero vev.
2) A Pati--Salam model where the Higgs field $H_{aIbJ}$ that couples to the
left sector is set to zero which is desirable because there is no symmetry
between the left and right sectors at low energies. 3) If one starts with
$\left(  D_{F}\right)  _{aI}^{\overset{.}{b}J}$ or $\left(  D_{F}\right)
_{aI}^{b^{\prime}J^{\prime}}$ or $\left(  D_{F}\right)  _{\overset{.}{a}%
I}^{\overset{.}{b}^{\prime}J^{\prime}}$ whose values are given by those that
were derived for the Standard Model, then the Higgs fields $\Sigma
_{aI}^{\overset{.}{b}J},$ $H_{aIbJ}$ and $H_{\overset{.}{a}I\overset{.}{b}J}$
will become composite and expressible in terms of more fundamental fields
$\Sigma_{I}^{J},$ $\Delta_{\overset{.}{a}J}$ and $\phi_{\overset{.}{a}}^{b}$ .
We refer to this as the composite model. 

Depending on the precise particle
content we determine the coefficients $b_{R},b_{L},b$ in \eqref{eq:RG-general}
that control the RG flow of the Pati--Salam gauge couplings $g_{R},g_{L},g$.
We run them to look for unification of the coupling $g_{R}=g_{L}=g$. The
boundary conditions are taken at the intermediate mass scale $\mu=m_{R}$ to be
the usual (e.g. \cite[Eq. (5.8.3)]{Moh86})
\begin{equation}
\frac{1}{g_{1}^{2}}=\frac{2}{3}\frac{1}{g^{2}}+\frac{1}{g_{R}^{2}},\qquad
\frac{1}{g_{2}^{2}}=\frac{1}{g_{L}^{2}},\qquad\frac{1}{g_{3}^{2}}=\frac
{1}{g^{2}},\label{eq:couplings-relations}%
\end{equation}
in terms of the Standard Model gauge couplings $g_{1},g_{2},g_{3}$. At the
mass scale $m_{R}$ the Pati--Salam symmetry is broken to that of the Standard
Model, and we take it to be the same scale that is present in the see-saw
mechanism. It should thus be of the order $10^{11}-10^{13}$Gev. We now discuss
the three models, in order of complexity.

\bigskip

\subsection{Pati--Salam with composite Higgs fields}

\begin{table}[ptb]%
\[%
\begin{array}
[c]{c|ccc}%
\text{particle} & SU(2)_{R} & SU(2)_{L} & SU(4)\\\hline
\phi_{\dot a }^{b} & 2 & 2 & 1\\
\Delta_{\dot a I} & 2 & 1 & 4\\\hline
\Sigma_{J}^{I} & 1 & 1 & 15
\end{array}
\]
\caption{Pati--Salam scalar particle content and their representations for a
first-order Dirac operator. The field $\Sigma_{J}^{I}$ in the last row is
decoupled if there is quark-lepton coupling unification.}%
\label{table:PS-part-cont}%
\end{table}We first consider the case of a finite Dirac operator for which the
Standard Model subalgebra $\mathbb{C}\oplus\mathbb{H}_{L}\oplus M_{3}%
(\mathbb{C})\subset\mathcal{A}_F$ satisfies the first-order condition
\cite{CC07b}. This condition is extremely constraining and forces the
couplings of the right-handed neutrino to be with a singlet. In this case, the
off-diagonal term in \eqref{eq:dirac} becomes
\begin{equation}
D_{\alpha I}^{\beta^{\prime}K^{\prime}}=\delta_{\alpha}^{\overset{.}{1}}%
\delta_{\overset{.}{1^{\prime}}}^{\beta^{\prime}}\delta_{I}^{1}\delta
_{1^{\prime}}^{K^{\prime}}k^{\ast\nu_{R}},
\end{equation}
and the diagonal structure of $D_{F}$ is determined by the following
sub-matrices \cite{CC10}
\begin{align}
D_{\alpha1}^{\quad\beta1} &  =\left(
\begin{array}
[c]{cc}%
0 & D_{a1}^{\overset{.}{b}1}\\
D_{\overset{.}{a}1}^{b1} & 0
\end{array}
\right)  ,\quad D_{a1}^{\overset{.}{b}1}=\left(  D_{\overset{.}{a}1}%
^{b1}\right)  ^{\ast}\equiv D_{a\left(  l\right)  }^{\overset{.}{b}}\\
D_{\alpha i}^{\quad\beta j} &  =\left(
\begin{array}
[c]{cc}%
0 & D_{a\left(  q\right)  }^{\overset{.}{b}}\delta_{i}^{j}\\
D_{\overset{.}{a}\left(  q\right)  }^{b}\delta_{i}^{j} & 0
\end{array}
\right)  ,\quad D_{\overset{.}{a}\left(  q\right)  }^{b}=\left(  D_{a\left(
q\right)  }^{\overset{.}{b}}\right)  ^{\ast}\nonumber
\end{align}
where
\[
D_{a\left(  q\right)  }^{\overset{.}{b}}=\left(
\begin{array}
[c]{cc}%
k^{\ast u} & 0\\
0 & k^{\ast d}%
\end{array}
\right)  .
\]
The Yukawa couplings $k^{\nu},$ $k^{e},$ $k^{u},$ $k^{d}$ are $3\times3$
matrices in generation space. Notice that this structure gives Dirac masses to
all the fermions, but Majorana masses only for the right-handed neutrinos.
One can also consider the special case of lepton and quark unification by
equating $k^{\nu}=k^{u},k^{e}=k^{d}$ which imply some simplifications.

The inner perturbations of the finite Dirac operator of the above type were
determined in \cite{CCS13b} to be composite fields $\Sigma_{\overset{.}{a}%
I}^{bJ}$ and $H_{\overset{.}{a}I\overset{.}{b}J}$, depending on fundamental
Higgs fields $\phi_{\overset{.}{a}}^{b}$, $\Sigma_{I}^{J}$ and $\Delta
_{\overset{.}{a}J}$ in the following way:
\begin{align}
\Sigma_{\overset{.}{a}I}^{bJ}  &  = \left(  k^{\nu}\phi_{\overset{.}{a}}%
^{b}+k^{e}\widetilde{\phi}_{\overset{.}{a}}^{b}\right)  \Sigma_{I}^{J}
+\left(  k^{u}\phi_{\overset{.}{a}}^{b}+k^{d}\widetilde{\phi
}_{\overset{.}{a}}^{b}\right)  \left(  \delta_{I}^{J}-\Sigma_{I}^{J}\right)
,\nonumber\label{dependent higgs}\\
H_{\overset{.}{a}I\overset{.}{b}J}  &  =k^{\ast\nu_{R}}\Delta_{\overset{.}{a}%
J}\Delta_{\overset{.}{b}I}.
\end{align}
The field $\widetilde{\phi}_{\overset{.}{a}}^{b}$ is not an independent field
and is given by%
\begin{equation}
\widetilde{\phi}_{\overset{.}{a}}^{b}=\sigma_{2}\overline{\phi}%
_{\overset{.}{a}}^{b}\sigma_{2}.
\end{equation}
We have listed the fundamental Higgs fields and their representations in Table
\ref{table:PS-part-cont}. We first assume that there is lepton quark
unification, so that the $\Sigma_{I}^{J}$ is decoupled.

The $\beta$-functions for the Pati--Salam couplings $g_{R},g_{L},g$ with the
above particle content are found to be
\begin{equation}
(b_{R},b_{L},b)=\left(  \frac{7}{3},3,\frac{31}{3}\right)  .
\end{equation}
The solutions of the RG-equations are easily found to be
\begin{align}
g_{R}(\mu)^{-2} &  =g_{R}(m_{R})^{-2}+\frac{1}{8\pi^{2}}\frac{7}{3}\log
\frac{\mu}{m_{R}},\\
g_{L}(\mu)^{-2} &  =g_{L}(m_{R})^{-2}+\frac{1}{8\pi^{2}}3\log\frac{\mu}{m_{R}%
},\\
g(\mu)^{-2} &  =g(m_{R})^{-2}+\frac{1}{8\pi^{2}}\frac{31}{3}\log\frac{\mu
}{m_{R}},
\end{align}
We impose the boundary conditions \eqref{eq:couplings-relations} at the mass
scale $\mu=m_{R}$.
\begin{figure}[ptb]
\includegraphics[scale=.25]{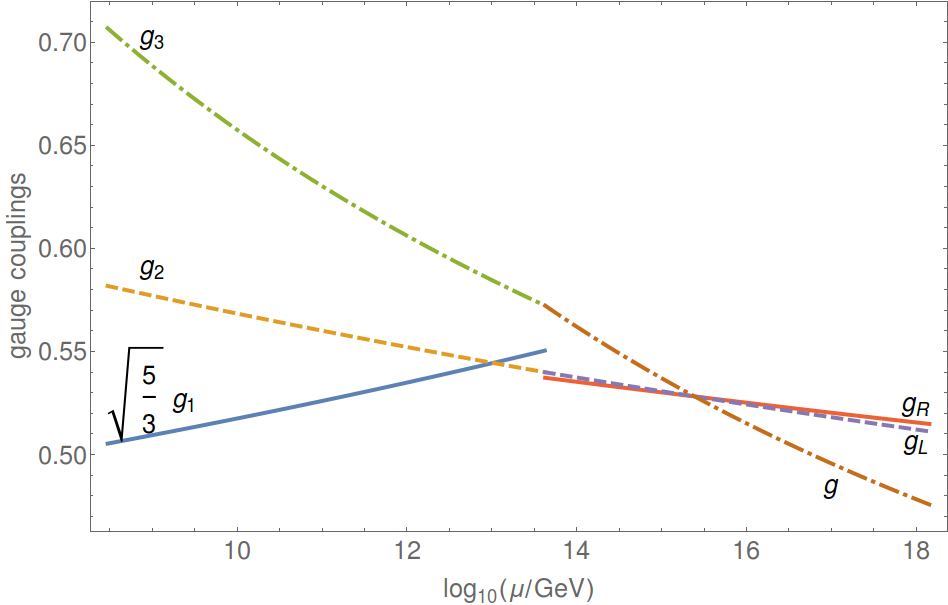}\caption{Running of coupling
constants for the spectral Pati--Salam model with composite Higgs fields:
$g_{1},g_{2},g_{3}$ for $\mu<m_{R}$ and $g_{R},g_{L},g$ for $\mu>m_{R}$ with
unification scale $\Lambda\approx2.5\times10^{15}\GeV$ for $m_{R}%
=4.25\times10^{13}\GeV$.}%
\label{fig:running-combined}%
\end{figure}As a first approximation, we adopt the usual running of the SM
gauge couplings, leaving a full analysis of the effect of the scalar fields
after Pati--Salam symmetry breaking to future work. Also, we ignore the
presence of non-renormalizable terms in the spectral action for the composite
model. Then, after experimenting with different values of $m_{R}$, we find a
unification scale $\Lambda\approx2.5\times10^{15}$Gev if we set $m_{R}%
=4.25\times10^{13}$Gev (Figure \ref{fig:running-combined}).

If the scalar field $\Sigma_{I}^{J}$ is not decoupled ---in other words, if
there is no lepton-quark coupling unification--- then there is an additional
scalar $(1_{R},1_{L},15)$ irreducible representation contributing to the
$\beta$-function, giving a slightly different $(b_{R},b_{L},b)=\left(
\frac{7}{3},3,9\right)  $. This in turn gives a unification scale
$\Lambda\approx6.3\times10^{15}$Gev for $m_{R}=4.1\times10^{13}$Gev.


\subsection{Pati--Salam with fundamental Higgs fields}

\begin{figure}[ptb]
\includegraphics[scale=.25]{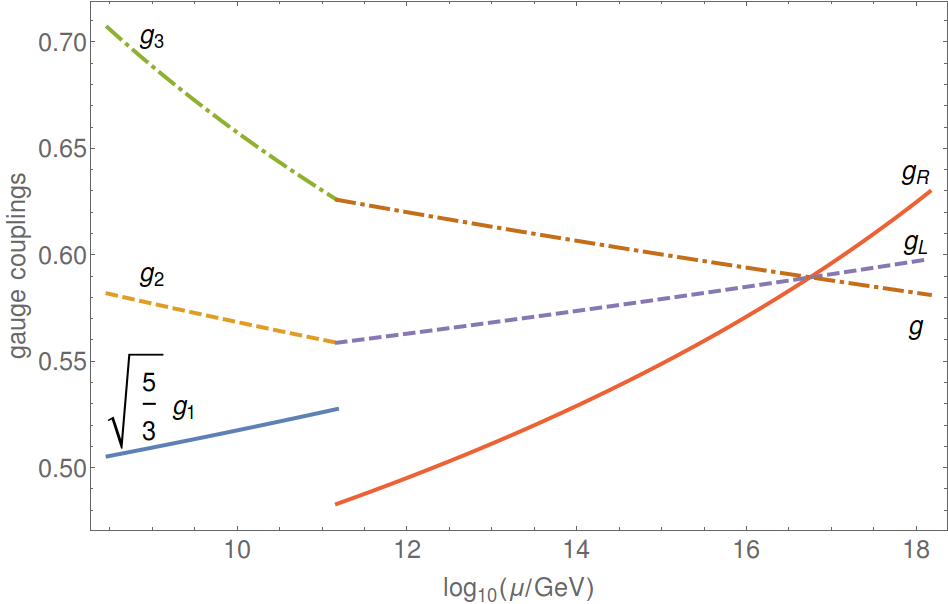}\caption{Running of
coupling constants for the spectral Pati--Salam model with fundamental Higgs
fields: $g_{1},g_{2},g_{3}$ for $\mu<m_{R}$ and $g_{R},g_{L},g$ for $\mu>
m_{R}$ with unification scale $\Lambda\approx6.3 \times10^{16} \GeV$ for
$m_{R} = 1.5 \times10^{11} \GeV$.}%
\label{fig:running-combined-NoOrder1}%
\end{figure}

Next, we consider the case of a more general finite Dirac operator, not
satisfying the first-order condition with respect to the Standard Model
subalgebra. We begin with the special case where%
\begin{equation}
\left(  D_{F}\right)  _{aI}^{b^{\prime}J^{\prime}}=0
\end{equation}
which implies that the Higgs field $H_{aIbJ}=0.$ The inner perturbations
$\Sigma_{\overset{.}{a}I}^{bJ}$ and $H_{\overset{.}{a}I\overset{.}{b}J}$ are
now themselves fundamental Higgs fields \cite[Sect. 5]{CCS13b} and their
representations are listed in Table \ref{table:PS-part-cont-NoOrder1}. The
$\beta$-functions are computed to be
\begin{equation}
(b_{R},b_{L},b)=\left(  -\frac{26}{3},-2,2\right)
\end{equation}
Note that the $SU(2)_{R}$ and $SU(2)_{L}$-sectors are not asymptotically free,
due to the large scalar sector. Nevertheless, we can still run the gauge
couplings with the boundary values set by \eqref{eq:couplings-relations}.
Adopting the same approximation as in the previous section, this results in
Figure \ref{fig:running-combined-NoOrder1}. The unification scale is
$\Lambda\approx6.3\times10^{16}$Gev if we set $m_{R}=1.5\times10^{11}$Gev.

\begin{table}[ptb]%
\[%
\begin{array}
[c]{c|ccc}%
\text{particle} & SU(2)_{R} & SU(2)_{L} & SU(4)\\\hline
\Sigma_{\dot a J}^{bJ} & 2 & 2 & 1+15\\
H_{\dot a I \dot bJ} \bigg\{ &
\begin{array}
[c]{cc}%
3 & \\
1 &
\end{array}
&
\begin{array}
[c]{cc}%
1 & \\
1 &
\end{array}
&
\begin{array}
[c]{cc}%
10 & \\
6 &
\end{array}
\end{array}
\]
\caption{Pati--Salam scalar particle content and their representations for a
general finite Dirac operator.}%
\label{table:PS-part-cont-NoOrder1}%
\end{table}

\subsection{Left-right symmetric Pati--Salam with fundamental Higgs fields}

As a final possibility we consider the most general case for $D_{F}$ which
gives in addition to the fundamental Higgs fields in Table
\ref{table:PS-part-cont-NoOrder1} the field $H_{aIbJ}$ in the $(1_{R}%
,3_{L},10)+(1_{R},1_{L},6)$ representation. The $\beta$-functions become
\begin{equation}
(b_{R},b_{L},b)=\left(  -\frac{26}{3},-\frac{26}{3},-\frac{4}{3}\right)
\end{equation}

\begin{figure}[ptb]
\includegraphics[scale=.25]{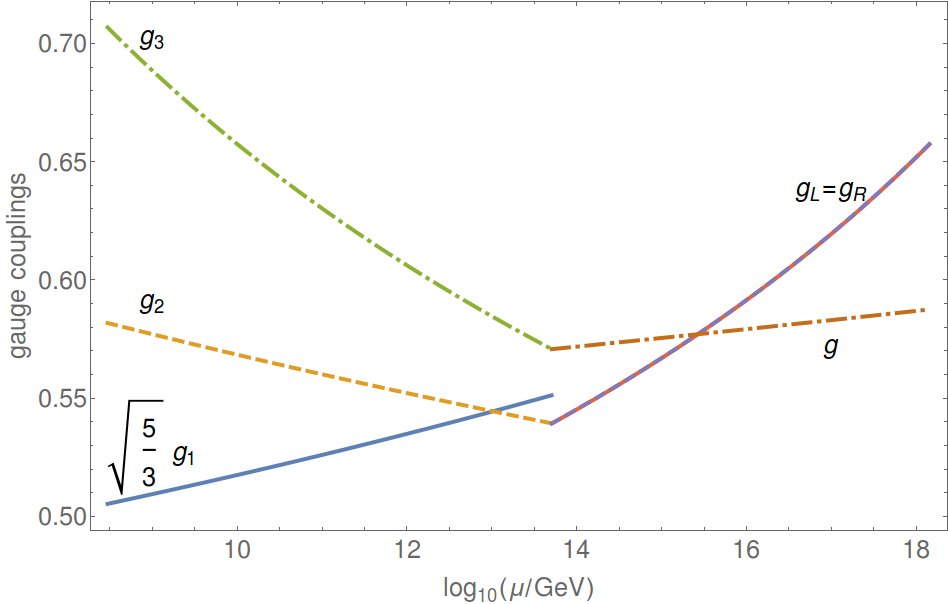}\caption{Running of
coupling constants for the left-right symmetric spectral Pati--Salam model:
$g_{1},g_{2},g_{3}$ for $\mu<m_{R}$ and $g_{R},g_{L},g$ for $\mu>m_{R}$ with
unification scale $\Lambda\approx2.7\times10^{15}\GeV$ for $m_{R}%
=5.1\times10^{13}\GeV$.}%
\label{fig:running-combined-NoOrder1LR}%
\end{figure}Adopting once more the approximation that we made use of in the
previous sections, we run the Pati--Salam gauge couplings from $m_{R}$,
resulting in Figure \ref{fig:running-combined-NoOrder1LR}. We find the
unification scale to be $\Lambda\approx2.7\times10^{15}$Gev if we set
$m_{R}=5.1\times10^{13}$Gev.

\section{Conclusions}

We have analyzed the running of the Pati--Salam gauge couplings for the
spectral model, considering different scalar field contents corresponding to
the assumptions made on the finite Dirac operator. We stress that the number
of possible models is quite restrictive and that one can not freely choose the
particle content. We have identified the three main models, although there
exists small variations on them. The different possibilities correspond to
restrictions on the geometry of the finite space $F$. In all the models
considered here, we establish unification of the gauge couplings, with boundary
conditions set by the usual Standard Model gauge couplings at an intermediate
mass scale.

Besides the direct physical interest of such grand unification, it also
determines the scale at which the asymptotic expansion of Equation
\eqref{eq:sa} is actually valid as an effective theory. In order to see this,
note that the scale-invariant term $F_{0}a_{4}$ in \eqref{eq:sa} for the
spectral Pati--Salam model contains the terms \cite{CCS13b}:
\begin{equation}
\frac{F_{0}}{2\pi^{2}}\int\left(  g_{L}^{2}\left(  W_{\mu\nu L}^{\alpha
}\right)  ^{2}+g_{R}^{2}\left(  W_{\mu\nu R}^{\alpha}\right)  ^{2}%
+g^{2}\left(  V_{\mu\nu}^{m}\right)  ^{2}\right)  . \label{eq:sa-ps}%
\end{equation}
Normalizing this to give the Yang--Mills Lagrangian demands
\begin{equation}
\frac{F_{0}}{2\pi^{2}}g_{L}^{2}=\frac{F_{0}}{2\pi^{2}}g_{R}^{2}=\frac{F_{0}%
}{2\pi^{2}}g^{2}=\frac{1}{4},
\end{equation}
which requires gauge coupling unification, $g_{R}=g_{L}=g$. Note that the
similar result for the Standard Model gauge couplings does not hold (at least
at the one-loop level) because the three couplings actually do not meet, even
though they are required to be unified in the spectral action \cite{CC96}. We
consider this to be strong evidence for the spectral Pati--Salam model as a
realistic possibility to go beyond the Standard Model.

To summarize, the spectral construction of particle physics models based on a
spectral triple with a noncommutative space with metric dimension four and
whose finite part has KO dimension $6$ leads directly to a family of
Pati--Salam models with gauge symmetry $SU\left(  2\right)  _{R}\times
SU\left(  2\right)  _{L}\times SU\left(  4\right)  $ and well defined Higgs
structure. Breaking of $SU\left(  2\right)  _{R}\times SU\left(  4\right)  $
to $U\left(  1\right)  \times SU\left(  3\right)  $ occurs at some scale
$m_{R}\sim10^{11}-10^{13}$ Gev with a unification scale where the three
coupling constants meet of the order of $10^{16}$ Gev$.$ All these breakings
will have the Standard Model as an effective theory at low energies.

\begin{acknowledgments}
The work of AHC is supported in part by the National Science Foundation
Phys-1202671. WDvS would like to thank Gert Heckman for pointing to reference
\cite{KP81}, and Florian Lyonnet for his help with PyR@TE. Also, WDvS thanks
IH\'{E}S for hospitality and support during a visit in June 2015 and NWO under
VIDI-grant 016.133.326.
\end{acknowledgments}

\end{document}